\documentclass{eas}
\usepackage{graphicx}

\def\etal{{et al.~}}
\def\Mo{{\rm M_\odot}}

\def\LCDM{$\Lambda$CDM}
\def\gtsim {\lower .1ex\hbox{\rlap{\raise .6ex\hbox{\hskip .3ex
        {\ifmmode{\scriptscriptstyle >}\else
                {$\scriptscriptstyle >$}\fi}}}
        \kern -.4ex{\ifmmode{\scriptscriptstyle \sim}\else
                {$\scriptscriptstyle\sim$}\fi}}}

\newcommand{\lsim}{\lower0.6ex\vbox{\hbox{$ \buildrel{\textstyle <}\over{\sim}\ $}}}
\newcommand{\gsim}{\lower0.6ex\vbox{\hbox{$ \buildrel{\textstyle >}\over{\sim}\ $}}}
\newcommand{\fcold}{f_{\mathrm{cold}}}

\begin{document}

\TitreGlobal{Mass Profiles and Shapes of Cosmological Structures}

\title{The Effect of Baryons on Halo Shapes}

\author{Stelios Kazantzidis}
\address{Institute for Theoretical Physics, University of Z\"urich,
CH-8057 Z\"urich, Switzerland.}
\secondaddress{Kavli Institute for Cosmological Physics and 
Department of Astronomy and Astrophysics, 
The University of Chicago, Chicago, IL 60637 USA.}
\author{Andrew R. Zentner}
\sameaddress{2}
\author{Daisuke Nagai}
\sameaddress{2}

\runningtitle{The Effect of Baryons on Halo Shapes}
\index{Stelios Kazantzidis}
\index{Andrew R. Zentner}
\index{Daisuke Nagai}

\begin{abstract} 

Observational evidence indicates a 
mismatch between the shapes of {\it collisionless} 
dark matter (DM) halos and those of observed systems.  
Using hydrodynamical cosmological simulations we 
investigate the effect of 
baryonic dissipation on halo shapes.  
We show that {\it dissipational} simulations 
produce significantly rounder halos than those 
formed in equivalent dissipationless simulations.  
Gas cooling causes an average increase in halo 
principal axis ratios of $\sim 0.2-0.4$ 
in the inner regions and a systematic shift 
that persists out to the virial radius, alleviating 
any tension between theory and observations.  
Although the magnitude of the effect may be 
overestimated due to overcooling, 
cluster formation simulations designed 
to reproduce the observed fraction of 
cold baryons still produce substantially 
rounder halos. Subhalos also exhibit a trend of increased axis ratios 
in dissipational simulations. Moreover, we demonstrate that subhalos are 
generally rounder than corresponding field halos 
even in dissipationless simulations. Lastly, we analyze a series of 
binary, equal-mass merger simulations of disk 
galaxies.  Collisionless mergers reveal a strong 
correlation between DM halo shape and stellar 
remnant morphology.  In dissipational mergers, 
the combination of strong gas inflows and star formation 
leads to an increase of the DM axis ratios in the remnant.  
All of these results highlight the vital role of baryonic 
processes in comparing theory with observations and warn against 
over-interpreting discrepancies with collisionless simulations 
on small scales.
\end{abstract}

\maketitle

\section{Introduction}

The hierarchical, cold dark matter (CDM) model 
of cosmological structure formation successfully 
explains a plethora of observations on large scales, 
where linear theory is adequate (e.g., Spergel \etal 2003).
However, on non-linear scales several issues remain 
unresolved.  One intriguing discrepancy concerns 
the shapes of dark matter (DM) halos.  
{\it Dissipationless} cosmological $N$-body simulations consistently 
produce halos with minor-to-major axis ratios 
in the range $\langle c/a \rangle \simeq 0.5 \pm 0.2$ 
and projected ellipticities of $\epsilon \sim 0.4-0.5$ 
(e.g., Jing \& Suto 2002, JS02).  
Inferring halo shapes observationally 
is a daunting task, yet a variety of probes exists 
that may enable us to distinguish between 
spherical and flattened halos (e.g., Merrifield 2004). 
In the Milky Way, flaring of the 
HI disk and the coherence of the Sagittarius 
tidal stream both imply a nearly spherical 
inner halo with $c/a \sim 0.7-0.9$ 
(e.g., Olling \& Merrifield 2000; Ibata \etal 2001).  
In other systems, the ellipticities of galaxies and clusters inferred 
from X-ray and optical isophotes (e.g., Buote \& Canizares 1996) 
and galaxy halos in weak lensing measurements 
(Hoekstra \etal 2004) are generally 
in the range $\epsilon \sim 0.2-0.3$.

Applying halo shapes as a cosmological test requires 
improvements in theoretical predictions.  
In this study, we investigate the effect of 
the dissipative infall of gas during galaxy formation 
on the shapes of halos and subhalos using high-resolution, 
cosmological simulations in the {\LCDM} cosmology.  
For the first time, we consider the effect of 
baryons on halo shapes in high-resolution gasdynamical 
binary merger simulations between 
multicomponent galaxies that include 
radiative cooling and star formation.

\section{Shapes of dark halos in cosmological simulations}

\begin{figure}[t]
   \centering
   \includegraphics[width=12cm]{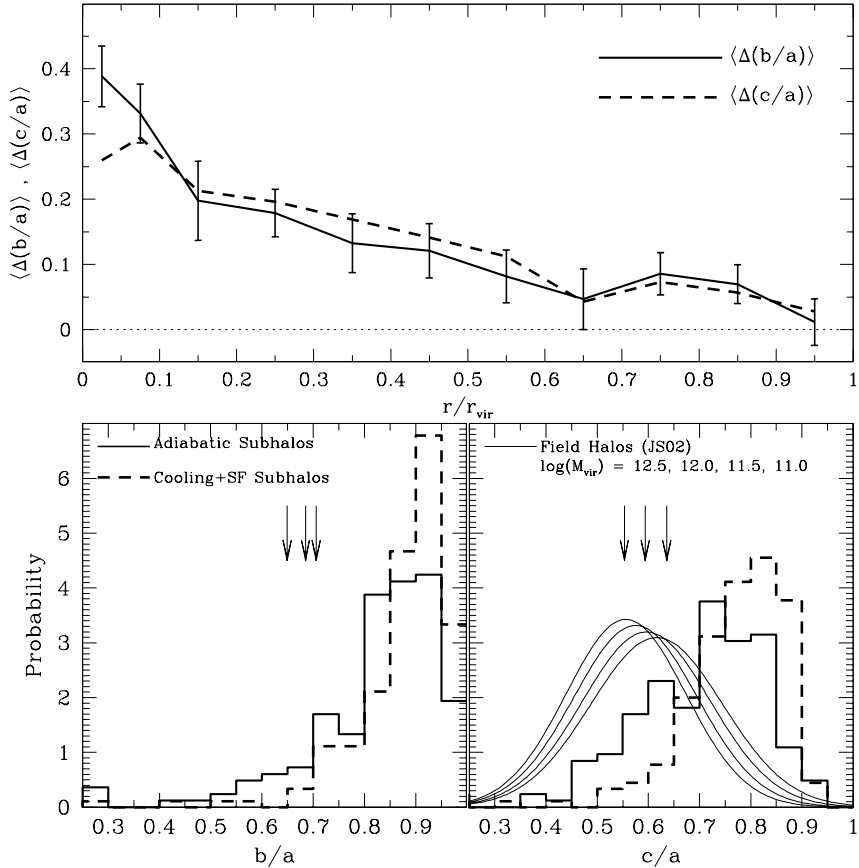}
      \caption{{\it Top panel:} The average difference between 
axis ratios in cooling and adiabatic cosmological simulations as a 
function of radius in units of the virial radius, $r_{\rm vir}$.  
The error bars correspond to the error on the mean value of 
$\Delta(b/a)$ in each bin.  The errors in $\Delta(c/a)$ are similar. 
{\it Bottom panels:} The distribution of axis ratios for subhalos in the 
simulated clusters. The {\it left} panel shows the distribution of 
intermediate-to-major axis ratio, $b/a$, and the {\it right} panel shows 
the minor-to-major axis ratio, $c/a$, distribution.  
{\it Solid} lines are for adiabatic simulations while 
{\it dashed} lines are for dissipational simulations.  
The {\it thin, solid} lines in the 
{\it right} panel show the distribution of $c/a$ for 
field halos using the fitting function of JS02, for 
four different masses indicated in the panel.  
The {\it downward arrows} show axis ratios of 
three host halos with $M_{\rm vir} \approx 10^{12}\  h^{-1}\Mo$ 
and with $\sim 10^6$ particles within their 
virial radii analyzed by Zentner \etal (2005).
\label{fig1}}
\end{figure}

In this section, we analyze cosmological 
simulations of $8$ cluster-size systems and one 
galaxy-size system in a flat {\LCDM} cosmology.  
For each system, we perform two sets of simulations 
starting from the same initial conditions 
but including different physical 
processes. The first set follows the dynamics 
of gas ``adiabatically'', without radiative cooling. 
The second set includes gas cooling, 
star formation, metal enrichment,
thermal supernovae feedback, 
and UV heating by an ionizing background.  
All cosmological simulations were performed with the 
Adaptive Refinement Tree code (Kravtsov 1999).  
For details on the simulations we refer the reader 
to Kazantzidis \etal (2004).  We determine 
principle axis ratios $b/a$ and $c/a$ ($a \ge b \ge c$) 
in differential radial bins using the modified inertia tensor 
(Dubinski \& Carlberg 1991) as described in detail in 
Kazantzidis \etal (2004).

Figure~\ref{fig1} summarizes our results.  
The top panel shows the average change in the axis
ratios of {\it DM} in cooling and adiabatic runs
[e.g., $\Delta(b/a) \equiv (b/a)_{\rm cool} - (b/a)_{\rm adiab}$]  
as a function of radius for all cosmological simulations.  
In the inner regions of halos, 
gas cooling increases axis ratios by 
$\sim 0.4$ at $0.1 r_{\rm vir}$ and by
$\sim 0.2-0.3$ at $0.3 r_{\rm vir}$. The difference decreases with radius, 
but persists out to the virial radius.
In the bottom panels, we present the 
distribution of {\it subhalo} shapes 
in the simulated clusters.  
The subhalo axis ratios we quote are 
based on particles within 
one half of the subhalo tidal radius, 
$r_{\rm t}$, and we analyzed only subhalos resolved with
$\ge 200$ particles within $r \le 0.5 r_{\rm t}$.  
The masses of subhalos selected by these criteria 
range roughly from $10^{11.3} - 10^{12.5}\  h^{-1}\Mo$.  
We show for comparison the shape distribution of 
collisionless {\it field} halos for similar masses from JS02 (solid line).  
First, subhalos in dissipationless cosmological simulations 
are considerably rounder than field halos, a result 
also reported by Moore \etal (2004).
Second, dissipational simulations predict rounder subhalos 
with a much more narrowly-peaked distribution of 
axis ratios.  

In the CDM paradigm, galaxies form from the 
condensation of baryons in halo centers. As the central baryonic 
concentration grows and forms a disk or a spherical system, the 
overall mass distribution becomes more centrally concentrated 
and the central potential becomes rounder. The box orbits of DM particles 
in the central parts that serve as the backbone of a triaxial 
mass distribution, adjust to the presence of this rounder potential and are 
changing into less elongated forms (e.g. Gerhard \& Binney 1985).
In a few crossing times, 
most of the remaining box orbits will be destroyed as the overall potential 
becomes rounder.  This transformation of orbital families drives 
halos to a more spherical shape, even at large radii, as our simulations 
demonstrate.  

\begin{figure}[t]
   \centering
   \includegraphics[width=8cm]{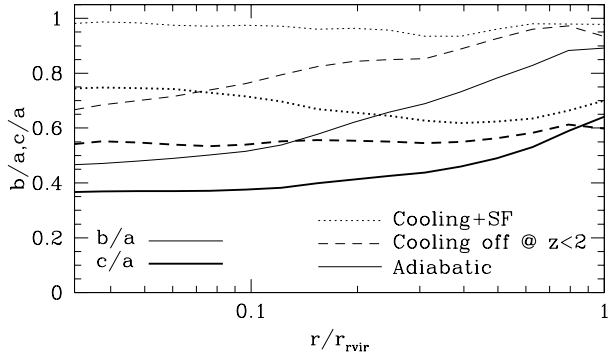}
      \caption{Principal axis ratios $b/a$ ({\it thin lines}) 
and $c/a$ ({{\it thick lines}}) of {\it DM} as a 
function of radius in units of $r_{\rm vir}$ for one of the Virgo-size clusters.  
{\it Solid} lines correspond to the adiabatic simulation, 
while the {\it dotted} lines show the results of the 
simulation with radiative cooling. The {\it dashed} lines show the DM shape 
profile in a simulation with cooling artificially turned off at $z \le 2$, 
resulting in a cold baryon fraction similar to that in observed 
galaxy clusters. 
\label{fig2}}
\end{figure}
%

\subsection{Halo shapes and the overcooling problem}

Current hydrodynamical cosmological simulations of clusters
may suffer from the ``overcooling'' problem:
the predicted fraction of cold baryons (stars and cold gas) 
$\fcold$, is a factor of $\sim 2 - 3$ higher than observed (Lin \etal 2003). 
Indeed, in our simulated clusters 
$\fcold \sim 0.3 - 0.4$, a factor of $\gsim 2$ 
higher than observed for systems in the mass range we 
consider.  Thus, the effect of dissipation on 
halo shapes may be overestimated. 

To assess the extent of the problem, we repeated one of the cluster 
simulations with radiative cooling {\it artificially} turned off at 
$z \le 2$. This simulation was designed to reproduce the observed value of 
$\fcold \approx 14\%$ at $z=0$.
Figure~\ref{fig2} shows the DM axis ratio profiles for the 
three different runs. The effect of dissipation in the simulation 
with radiative cooling turned off is smaller 
than in the original simulation, as fewer baryons condense in the center,
yet the shape of the DM distribution is 
still considerably more spherical than in the adiabatic run.
This result suggests that the effect of gas cooling on halo shapes 
is substantial even in simulations with realistic cold baryon fractions.

\section{Shapes of dark halos in merger simulations}

In this section, we analyze high-resolution simulations 
of binary equal-mass mergers
of disk galaxies including different physical processes.
We simulate purely collisionless 
mergers and mergers in which we follow the gas dynamics adiabatically.  
A third set of simulations includes radiative cooling, 
while in a fourth set we also incorporate
star formation.  We construct galaxy models using the 
technique of Hernquist (1993) 
with structural parameters consistent with standard 
{\LCDM} (Mo \etal 1998).  
Each galaxy consists of a spherical and isotropic Navarro \etal (1996) 
DM halo and an exponential stellar disk, 
and may contain a spherical, Hernquist (1990) bulge and/or an exponential 
gaseous disk.  
The dark halo mass and total disk mass are constant in all initial galaxy models. 
We considered collisions on parabolic orbits 
with two encounter geometries 
(randomly-inclined and coplanar disk orientations), 
two values for the gas fraction 
(10\% and 50\% of the total disk mass), 
and two values for the disk thickness.  
All merger simulations were performed with the GASOLINE TreeSPH $N$-body
code (Wadsley \etal 2004).
For further details on the merger simulations we refer the reader to 
Kazantzidis \etal (2004) and Kazantzidis \etal (2005).

\begin{figure}[t]
   \centering
   \includegraphics[width=12cm]{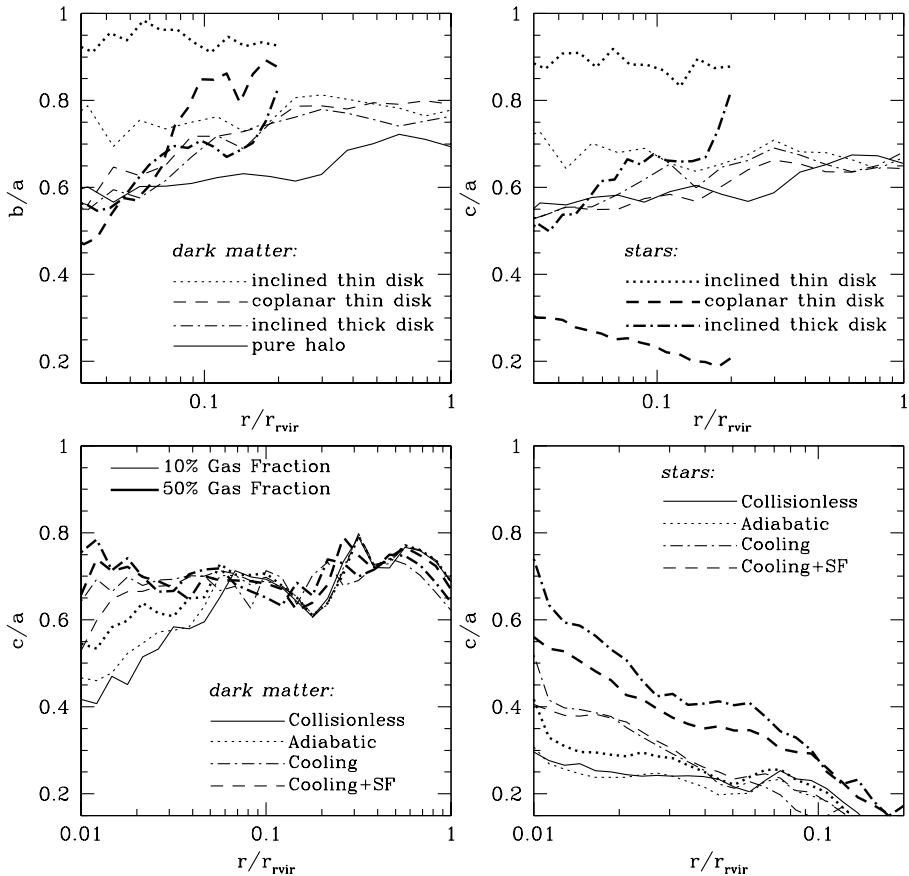}
      \caption{ 
{\it Top Panels:} Remnant axis ratio profiles in {\it collisionless}, 
equal-mass binary mergers of disk galaxies as a function of radius in units of the virial radius 
of the remnant, $r_{\rm vir}$. {\it Thin} lines show axis ratios for DM, 
while thick lines show results for stars. The orientation and thickness of the disks 
are indicated in each panel. 
{\it Bottom Panels:}  Remnant axis ratio profiles in collisionless and hydrodynamical, 
equal-mass binary disk galaxy mergers in the coplanar encounter geometry. 
The {\it left} panel shows results for the DM, while the {\it right} panel shows results for 
the stellar distributions. {\it Thin} and {\it thick} lines correspond to simulations with gas 
fractions of $10\%$ and $50\%$ in the initial merging disks, respectively.  
Different physical processes are included in each merger as indicated in the panels.
}
\label{fig3}
\end{figure}

In the top panels of Figure~\ref{fig3}, we present radial profiles $b/a$ and 
$c/a$ for remnants in {\it collisionless}, equal-mass binary merger simulations of halos
with embedded stellar disks.  For comparison we also
show results of a DM halo-only merger.  
Inclined disk mergers produce nearly spherical stellar 
remnants, while coplanar mergers yield a disk-like stellar remnant 
with $c/a\sim 0.3$.
The corresponding DM distribution is also more spherical 
in the inclined disk merger. However, in an inclined disk merger 
with the same orbital parameters as before, but with a factor of three thicker 
stellar disks, we find that the stellar remnant 
becomes less spherical and the halo axis 
ratios are lower by $\sim 0.15$ in the inner regions. 
These results indicate that the shape of the stellar remnant is 
sensitive to the mutual orientation and internal 
properties of the merging disks. The presence of a stellar remnant modifies the 
potential in the central region, where stars dominate gravitationally, and 
consequently affects the shape of the DM distribution.

The lower panels of Figure~\ref{fig3} show 
results for collisionless and hydrodynamical equal-mass, binary
merger simulations of disk galaxies containing bulges. 
Dissipational mergers lead to an 
increase in axis ratios of DM that sensitively depends on the details of gas physics 
included in the simulations. The largest increase occurs in dissipational simulations 
which exhibit the deepest central potentials.  
In the adiabatic simulations, axis ratios are close to 
those of the collisionless runs because the gas is too hot 
($T >10^5$~K) to build up a significant central mass concentration.
Since the DM responds to the {\it total} baryonic density enhancement
DM axis ratios are similar in all dissipational mergers.  

\section{Conclusions}

The dissipative infall of gas during galaxy formation alleviates 
the discrepancy between the shape distribution of collisionless 
DM halos and that of observed systems. 
Using high-resolution cosmological simulations we show that 
halos in simulations with radiative cooling are considerably 
more spherical than those in dissipationless simulations.
Dissipation results in an average increase in principal axis 
ratios of halos by $\sim 0.2-0.4$ in the inner regions.  
The difference decreases slowly with radius but remains substantial 
even at the virial radius.  Subhalos also exhibit a similar trend of increased axis
ratios in dissipational simulations and are generally rounder than corresponding 
field halos. We also showed that in an {\it ad hoc} cluster formation simulation
designed to reproduce the observed cold baryon fraction, the DM halo 
is still notably rounder at all radii compared to its dissipationless 
counterpart, suggesting that our results are robust despite any 
issues of overcooling in simulated clusters.

In hierarchical CDM models of structure formation, halos grow 
during periods of slow accretion, when gas cools and condenses toward the center, 
and sequences of violent mergers.
We performed a comprehensive series of collisionless and gasdynamical binary merger 
simulations between 
disk galaxies to study the effect of baryons on the shape 
of the remnant DM distributions.  
Collisionless simulations indicate a 
strong correlation between the shape of the 
stellar remnant and DM halo shape.  
In dissipational mergers, 
strong gas inflows and star formation drive the remnant DM distribution 
to be significantly more spherical. All of the results reported in this study underscore the 
crucial role of baryonic processes in comparing the small scale predictions of the {\LCDM} 
paradigm with observations.

\section*{Acknowledgments}

We would like to thank our collaborators Brandon Allgood, Monica Colpi, 
Victor Debattista, Andrey Kravtsov, Piero Madau, Ben Moore, 
Tom Quinn, Joachim Stadel and James Wadsley for allowing us 
to present our results here.  We also wish to thank the organizers of the 
XXIth IAP conference ``Mass Profiles \& Shapes of Cosmological 
Structures'' for a stimulating meeting which motivated part of this work.
SK is supported by The Kavli Institute for Cosmological Physics (KICP) 
at The University of Chicago. ARZ is funded by the KICP and The National 
Science Foundation under grant No. NSF PHY 0114422.
DN is supported by the NASA Graduate Student Researchers 
Program.  The cosmological simulations were performed on the 
IBM RS/6000 SP4 system at the National Center for Supercomputing Applications.
The merger simulations were performed on the zBox 
supercomputer at the University of Z\"urich and on the 
Intel cluster at the Cineca Supercomputing Center in Bologna.

\end{document}